\documentstyle[preprint,prb,aps,epsf]{revtex}

\begin{document}
\begin{center}
{\Large {\bf  Self - assembly of HgTe nanoparticles into nanostars using
    single stranded DNA }} \\ 
{\bf Satchidananda Rath and Surendra Nath Sahu\footnote{{Corresponding
      author}\\ 
{E-mail:sahu@iopb.res.in}}}\\
{\it {{} Institute of Physics, Sachivalaya Marg,
Bhubaneswar - 751 005, India.}}\\
\end{center}
\begin{abstract}
Self-assembly of structurally tunned nearly monodispersed HgTe - ssDNA
nanostars of average size 1.4 nm have been achieved by
manipulating HgTe nanoparticles using single stranded (ss) DNA  under 
galvanostatic condition. The nanostars are linked with one another by 
ssDNA of length 0.35 nm. Whereas, HgTe nanoparticles without ssDNA
complexation show a polycrystalline character with size in the range
4 - 7 nm. Room temperature photoluminescence (PL) of HgTe - 
ssDNA nanostars have yielded a single narrow PL at 548.4 nm of width 8 nm that 
corroborates to the nearly monodispersity of the nanostars and
predicts the lateral exciton transfer. On the other hand, 
polydispersed HgTe nanoparticles exhibit free- and bound- exciton dominated
luminescence in their PL spectrum.\\ 
\end{abstract}

\noindent
Self-assembly of semiconductor particles and clusters in the nanoscale 
has become increasingly important in nanoscale science and
technology for their novel properties derived from their size 
quantization\cite{gapo,sahu,arn,apa} and surface
effects\cite{apa,kkn}. Assembly of these nanostructures by conventional 
technique have yielded polydispersity and have encountered difficulties 
for the evaluation of their optical properties and even limited to their 
nanoscale device applications\cite{patel,kkn2}. Thus, a
bottom-up approach like wet chemistry or electrochemistry can be
employed to synthesize nanostructured semiconductors with a good
control over the size which is uncommon at this point of time. Assembly of
such nanoparticles by sequence selective 
DNA could be a good strategy to achieve monodispersity
and ordered nanoparticles. 
DNA is an important constitutent of the biological system consisting 
of nitrogenous bases like Purine (Guanine, G and Adenine, A) and
Pyrimidine (Cytosine, C and Thyamine, T), deoxyribose sugar and
phosphate group as an unit block.  
These are dictated by a network of specific hydrogen bonding interaction 
involving purine and pyrimidine bases with base to base separation as 3.4 
$ A^{o}$. Because of the presence of phosphate group, the backbone is
negatively charged, giving polyelectrolyte (charges of same sign bound
to a specific site of a polymer) nature to it.
An interesting property associated with DNA molecule is it's elctrostatic
interaction with positively charged ions, cations\cite{ker,gel} in
the solution analogous to 
the biological systems relevent to the DNA within a cell\cite{abe}. In
particular, to organize DNA in chromatin (in eukariotic cells), nature utilize
proteins having large positive charges histones and can form a stable
bead-on-a- string nekless structure. Hence, it should be possible that 
such electrostatic interaction can self-assemble the nanocrystals
under an electric bias onto a conducting substrate. 
In the present work we describe a new and
simple strategy to synthesize mercury telluride (HgTe) - single
stranded (ss) DNA nanostars taking above interaction of ssDNA into account and
show the emergence of different structure, morphology and optical properties in
comparision to HgTe nanoparticles without ssDNA. The unique optical properties
involving lateral inter-nanostar exciton transfer as a result of self -
assembling of nanoparticles in nanostar using ssDNA is being demonstrated 
by the photoluminescence measurement.\\

Although different approaches to assemble nanoparticle - DNA complex 
have been described in the literature\cite{chad,wang,mae,mae1}, most
of them have concentrated on  colloidal particles (metallic nanoparticles) 
aggregation process based on light scattering and transmission electron 
microgaph studies. None of them have
considered the structural and/or photoluminescence (PL) aspects where
the later is more pertinant to biosensor/identifying genetic disorder
applications. For biosensors, one should have a very narrow PL line width
which would be possible only from monodispersed nanocrystal
semiconductors. Such an attempt has not been explored so far. In our
approach, we have taken commercially available ssDNA
$\acute{5}GCAAGCGGTGAACCAGTTGTG\acute{3}$ of 7.5 nm length with 21
base. $Hg^{2+} (0.014 M)$ and $Te^{4+} (0.026 M)$ were made available from 
their respective salt solutions. The electrolytic bath containing
ssDNA, $Hg^{2+}$ and $Te^{4+}$ ions had pH = 0.6 at 278 K. The
electrodes were indium tin oxide (ITO) coated conducting glass -
cathode and Platinum - anode. Electrodeposition under stirring
condition was carried out at 1.5 $mA/cm^{2}$ cathodic current for 2
minute. Although at 278 K, the ssDNA can form hairpin structure,
in the stirring environment, there is always a restoring force
acting on the ssDNA. Hence, for a guide to the eye, it has been 
represented by a coiling structure as shown in Fig. 1. Short chain ssDNA
was choosen to avoid unwanted breaking of the chain during stirring
and to get finite and separated HgTe - ssDNA complex nanostars.
To check the stability of ssDNA in the working environment, optical absorption
measurements were performed for 0.3 OD ssDNA in milipore water at pH= 0.6 and
278 K temperature. No appreciable change in absorption spectrum
was noticed which confirms it's stability.
In order to show superiority of our approach, we have also
grown HgTe nanoparticles without ssDNA complexation keeping the deposition
parameters same as described above. 
Our approach to synthesize HgTe - ssDNA complex nanostars is schematically 
outlined in Fig. 1. In step (1), the ssDNA is in the solution containing 
$Hg^{2+}$ and $Te^{4+}$ ions under continuous stirring environment. The 
negative potential due to charge distribution in ${P{O_{4}}^{2-}}$ group of
ssDNA is largely 
compensated by the positive potential due to counter ions (positively
charged) and condensation of counter ions start taking place with non- 
specific binding governed mainly by coulomb attraction. By that the
ssDNA experiences an increase  
in entropy due to charge inversion\cite{gro} and starts wrapping
around counter ions (in this case $Hg^{2+}$ and $Te^{4+}$) to satisfy 
the charge
neutrality condition as shown in step (2). In step (3), upon
impression of a current/potential between two electrodes, the ssDNA
bases (G and A) are oxidized due to low oxidation potential\cite{lew}. As
a result, partial charges are developed near G and A which generate a
complex electric field whose distribution in the solution is given by
$xA + yB = z(A + B)$ where
A $\rightarrow$ Purine, B$ \rightarrow$ Pyrimidine, $ x= x_{1} +
x_{2}$ and $y= y_{1} + y_{2}$, $x_{1}$ and $x_{2}$ are the total
partial positive and negative 
charge developed near A and $ y_{1}$ and $y_{2}$ are the total
partial positive and negative charge developed near B where 
z is the total partial charge developed near the ssDNA bases. 
The coulomb interaction between the charges  play an important role
both in the formation of the structures and in transport processes.
Thus, the $Hg^{2+}$ and $Te^{4+}$ ions are
wrapped by the ssDNA after a charge inversion process which results
in the generation of excess positive charge and form a complex with
ssDNA in a random fashion. Upon impression of an electric
field between the two electrodes the $Hg^{2+}$ $\&$ $Te^{4+}$ - ssDNA
complex is alligned and dragged towards the cathode. The wrapped
$Hg^{2+}$ and $Te^{4+}$ ions now take the required number of electrons from the
cathode and subsequently hybridize themselves to form HgTe - ssDNA complex 
nanostars on the cathode\cite{pand}.\\ 

As there are small number of HgTe nanoparticles in the
wrapped DNA molecule, one would expect the formation of single
crystals HgTe - ssDNA complex.
Fig. 2 shows the TEM micrograph and diffraction pattern (inset) of HgTe
without  ssDNA taken at 80 KeV electron beam energy. The size of HgTe
crystallites are in the range of 4 - 7 nm and
non-spherical. The electron diffraction patterns clearly depicts
polycrystalinity with preferred orientation in the sample and has the
cubic structure. On the other hand the TEM micrograph and diffraction
pattern (inset) shown in Fig. 3 depicts the HgTe - ssDNA complex
nanostars are organised in a finite order with
uniform in their lateral size (1.4 nm) and nanostars are linked 
with one others by ssDNA of length around 0.35 nm, incidentally this
is also roughly the base to base separation of ssDNA (3.4 $A^{o}$). Our
ssDNA length is 7.5 nm and crystalline size 
is 1.4 nm. If we assume  that the ssDNA makes two and half turns on
each HgTe nanocrystals (because of charge inversion) then, in that
case, the ssDNA  can entangle at
best two HgTe nanoparticles and form the nanostars as observed from TEM
micrograph. The rest 0.5 nm ssDNA may have linked to another
crystal or form a tentacle. Fig. 3 inset shows the diffraction pattern of
HgTe - ssDNA deposit. Surprisingly the feature shows monocrystal
structure and has cubic phase. The present work clearly demonstrates a superior
self-assembling process for the HgTe - DNA systems.

The HgTe nanoparticles without ssDNA have the size in the range 4-7 
nm which is much less than the Bohr exciton radius (40 nm). So, a prominent 
strong quantum confinement effect (QCE) has been observed with these
nanocrystals.  Thus, if such nanoparticles are excited with high energy
photons (say UV light), one would expect light emission\cite{kkn1} in the
visible regime which can help in biosensor or electroluminescence device
applications. Indeed, our HgTe nanoparticles emit
yellow-orange light at room temperature when excited with UV radiation 
($\lambda$ = 369 nm). 
Fig. 4 (a) is the photoluminescence spectrum of HgTe
nanoparticles which consists two peaks, $P_{1}$ and $P_{2}$ peaking at 
579.5 nm and 588.3 nm with full width of half 
maxima (FWHM) as 35 nm and 26 nm respectively and it needs
explanation. Since localization energy of the exciton is the energy
difference between peaks $P_{1}$ and $P_{2}$ which is around 31 meV and
energy corresponding to room temperature is 26 meV and are comparable
hence, some of the 
excitons are thermally released from the bound state as
free exciton. As a result the luminescence, $P_{1}$
corresponding to free exciton transition and the red
shifted narrow band $P_{2}$ whose energy is less than $P_{1}$ can be
ascribed to the bound exciton
transition. The narrowness of bound-exciton PL is due to the large
spatial overlapping of initial and final state wave functions resulting 
increased in oscillator strength and short lifetime than free-exciton PL.
If one assumes the effective mass
approximation (EMA) then the average size estimated from the band edge PL
(average size is 5.14 nm) does not match with TEM measurement. Such a
descripancy is attributed to the non-sphericity in the nanoparticle and
the particle in a finite potential well. On the contrary, the  HgTe -
ssDNA nanostars PL spectrum shown  
in Fig. 4 (b) gives a very narrow single peak at 548.4 nm with FWHM as 
low as 8 nm. Origin of blue shifted narrow PL band compared to Fig. 4
(a) can be understood using the following 
reasons: (i) localization of the excitons as a result of carrier
confinement by the ssDNA boundary of the nanostars acting as a
resonator\cite{gam,bon}, (ii) if the separation 
between two quantum dots (QD) is small enough than the
exciton dimension, then transfer of exciton from one QD to another is possible
through diffusion mechanism, as a result of which the excitons of
different QDs are populated and undergo coherent radiative
recombination resulting a very narrow PL band \cite{rob} . Since the nanostars
are linked with each other by ssDNA of length 0.35 nm as observed from
TEM measurement hence, 
interaction between self - assembled nanoparticles through diffusion
mechanism have resulted population of excitons which ungergo coherent
radiative recombination to yield such a narrow PL line width.
Such a narrow single PL peak suggests a nice
perfection in the HgTe - ssDNA complex nanostars and clear observation 
of lateral exciton transfer between self - assembled nanoparticles.
Such a single sharp PL peak have tremendous application in biosensors by
anchoring the same with the conjugate ssDNA of
$\acute{5}GCAAGCGGTGAACCAGTTGTG\acute{3}$. A very narrow PL peak is a 
prerequisite for any biosensor application if one intends to use PL as 
an optical probe for biosensor applications. The nanostars may even
lead to identification of genetic disorder.\\

In summary, the present work clearly demonstrate an approach to use DNA
molecules for complexation of precursor cations( $Hg^{2+}$ and
$Te^{4+}$), followed by their 
transport to and electrochemical reduction on the cathode and
synthesize a new class of HgTe - ssDNA complex nanostar
hybrid structures. Such structures are easily
controllable and have resulted narrow line width luminescence 
and structural properties that are far different than HgTe
nanoparticles. The most important spectral evidence of lateral excitons
transfer is observed from PL measurement as a result of self -
assembling of the nanoparticles into nanostars by ssDNA which have resulted
narrow PL line width and can be used as a material candidate for
biosensor application.\\

Thanks are due to Prof S.N. Behera, Prof. R.K. Choudhury,
Institute of Physics, Bhubaneswar, India, Dr. K. K. Nanda, University
of Dublin, Ireland, for their helpful discussion 
and constant encouragement. Dr. M. Vijaylakshmi, IGCAR Kalpakkam, India, is
acknowledged for helping in TEM measurement.The author would like
to acknowledge Mr. S. N. Sarangi for his help during experiments.\\

\newpage
\noindent {\bf Figures caption}\\

Fig. 1. Schematic representation of the mechanism taking place during 
 synthesis of HgTe - ssDNA complex nanostars.\\

Fig. 2. TEM micrograph of HgTe nanoparticles, inset the diffraction
pattern.\\

Fig. 3. TEM micrograph of  HgTe - ssDNA complex nanostars, inset the
  diffraction pattern.\\ 

Fig. 4.  PL spectrum of HgTe nanoparticles (a) and HgTe - ssDNA complex
  nanostars (b). Solid line is the gaussian fitted curve of the PL spectrum.\\
\newpage
\begin{figure}[htbp]
\protect\centerline{\epsfxsize=5in \epsfbox{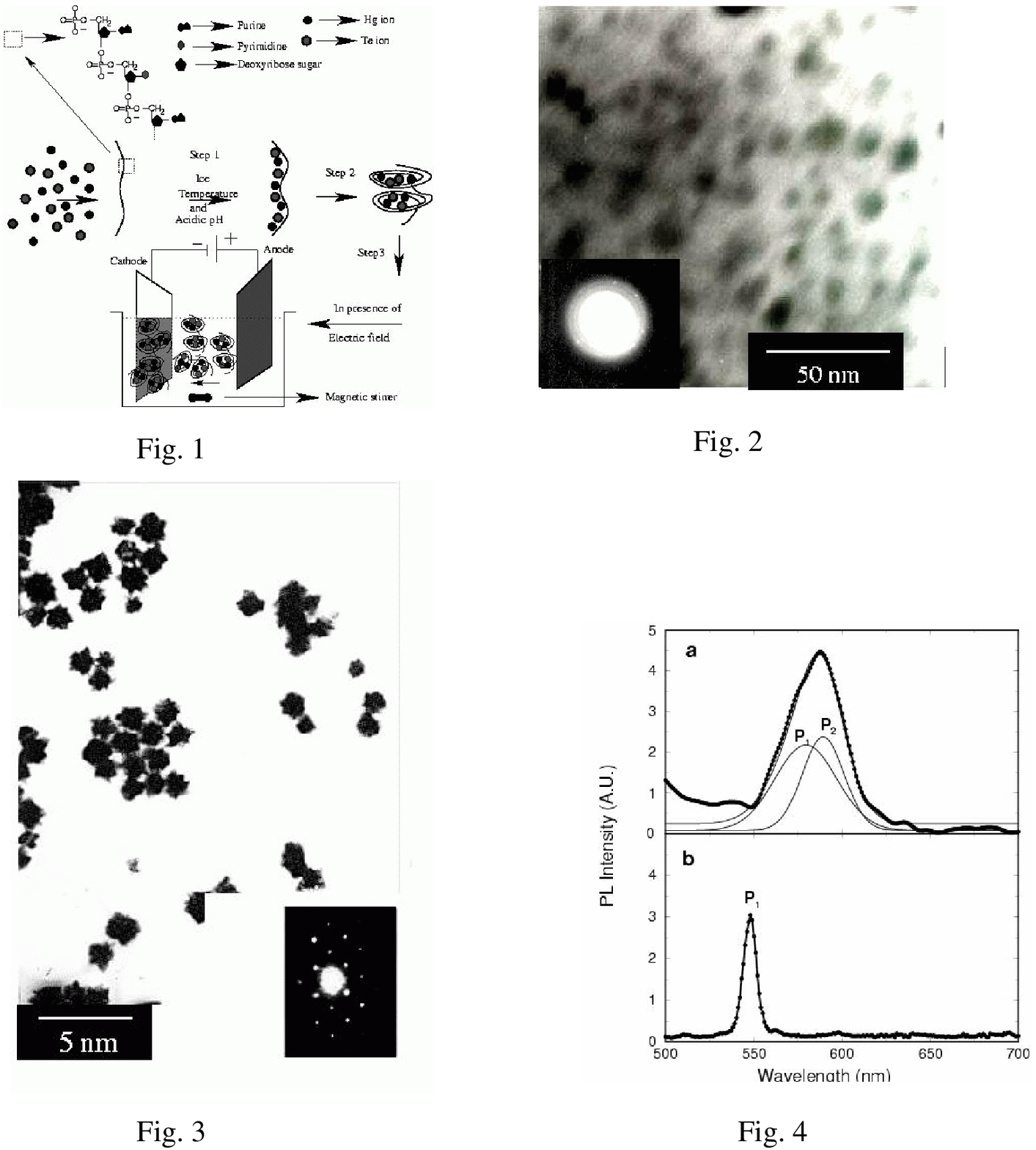}}
\end{figure}
\end{document}